\documentclass[tog]{acmsiggraph}
\usepackage{algorithm2e}

\def\BibTeX{{\rm B\kern-.05em{\sc i\kern-.025em b}\kern-.08em
    T\kern-.1667em\lower.7ex\hbox{E}\kern-.125emX}}

\TOGonlineid{45678}

\TOGvolume{0}
\TOGnumber{0}

\title{Robust Watertight 2-Manifold Surface Generation Method for ShapeNet Models}

\author{Jingwei Huang\thanks{e-mail:jingweih@stanford.edu} , Hao Su\thanks{e-mail:haosu.work@gmail.com}, Leo Guibas\thanks{e-mail:guibas@cs.stanford.edu}}


\begin{document}


 \teaser{
   \includegraphics[width=0.4\textwidth]{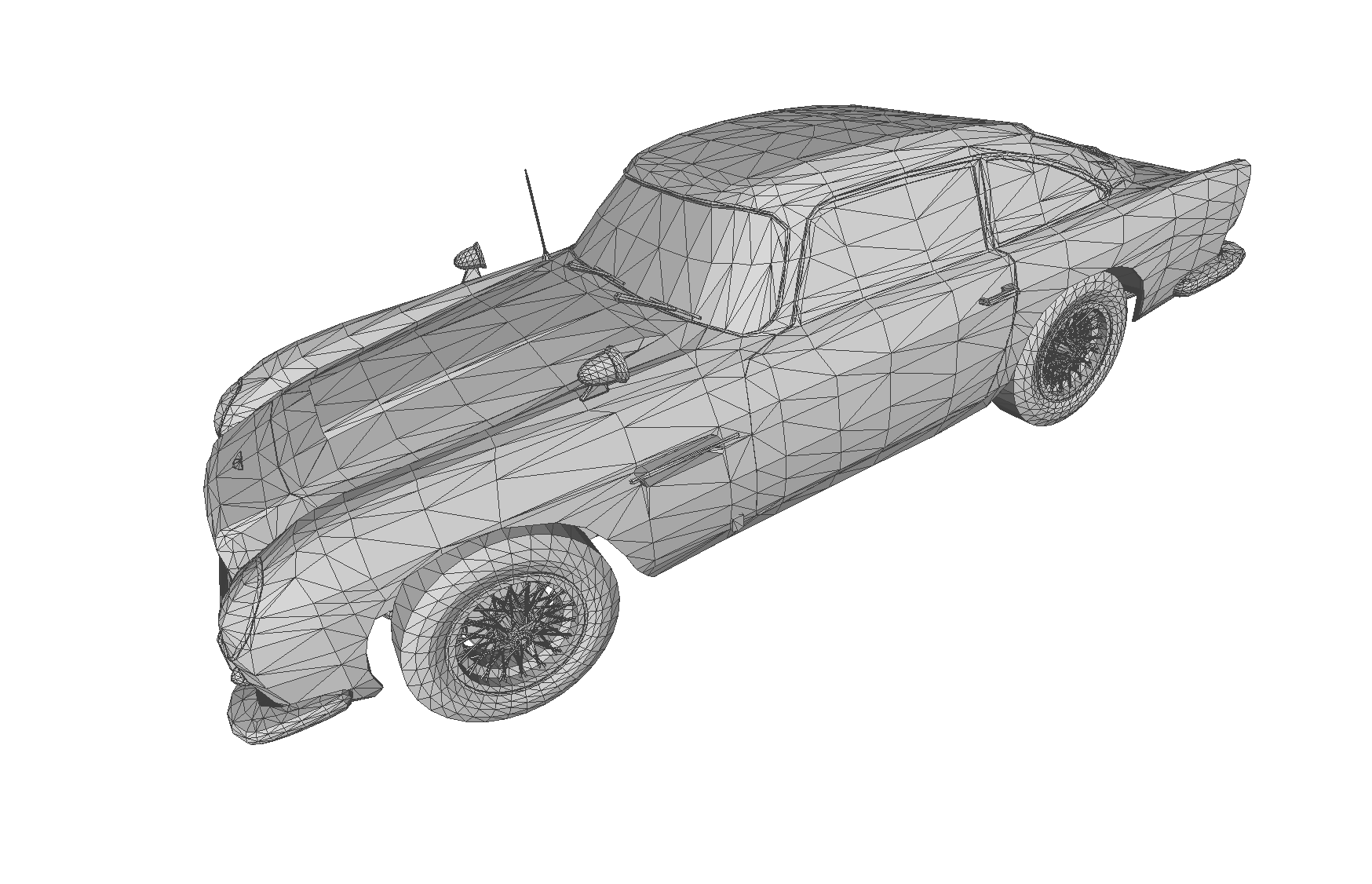}
   \includegraphics[width=0.4\textwidth]{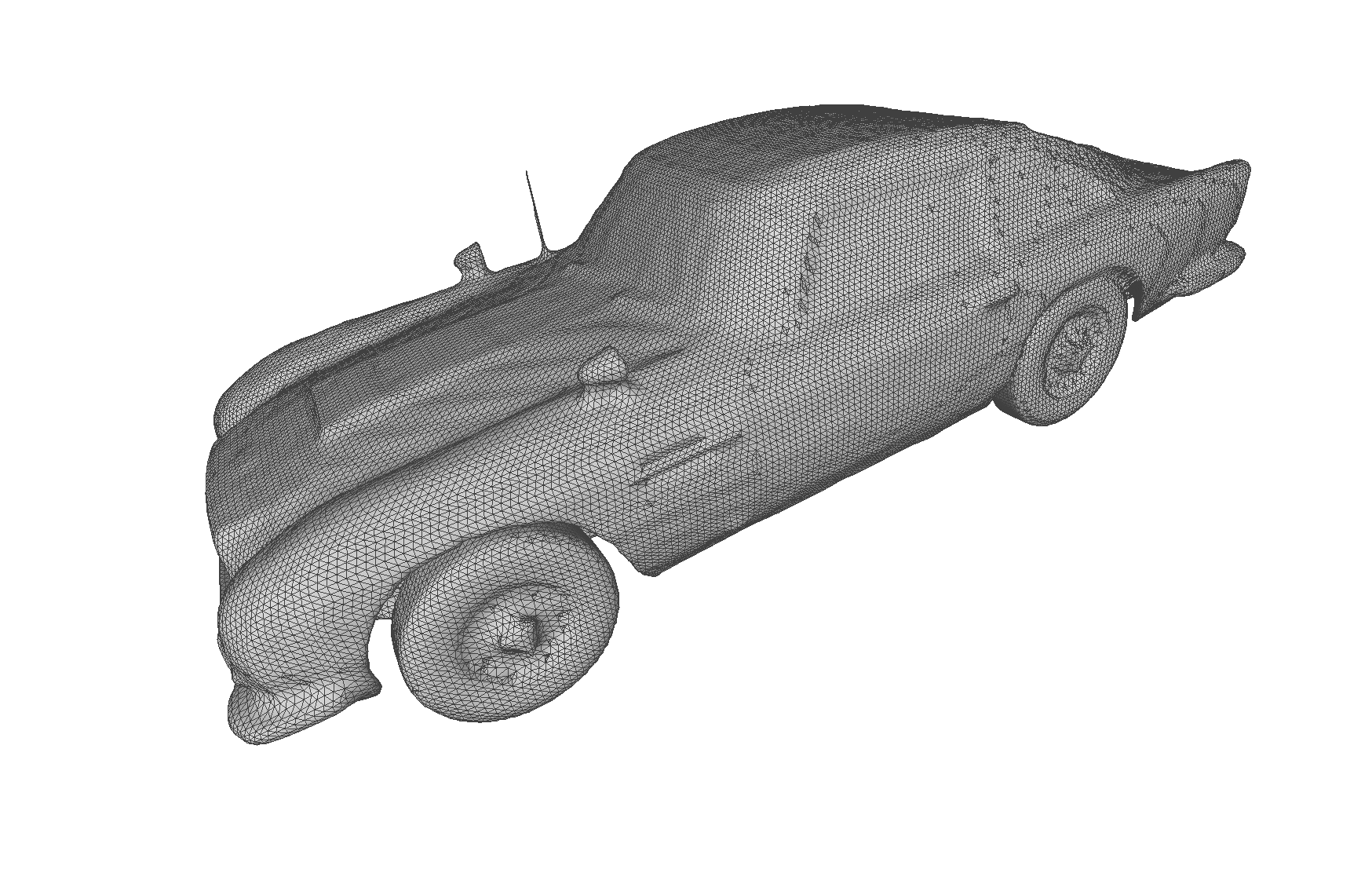}
    \caption{2-Manifold Processing}
 }

\maketitle

\begin{abstract}

In this paper, we describe a robust algorithm for 2-Manifold generation of various kinds of ShapeNet Models. The input of our pipeline is a triangle mesh, with a set of vertices and triangular faces. The output of our pipeline is a 2-Manifold with vertices roughly uniformly distributed on the geometry surface. Our algorithm uses an octree~\cite{meagher1982geometric} to represent the original mesh, and construct the surface by isosurface extraction. Finally, we project the vertices to the original mesh to achieve high precision. As a result, our method can be adopted efficiently to all ShapeNet models with the guarantee of correct 2-Manifold topology.

\end{abstract}

\begin{CRcatlist}
\CRcat{I.3.5}{Computer Graphics}{Computational Geometry};
\end{CRcatlist}

\keywordlist


\copyrightspace

\section{Data Presentation}

Our method for manifold surface generation is based on isosurface extraction. Therefore, the first step of our algorithm is to convert the input triangle mesh into the signed distance field. The simplest method is to use a uniform grid to store the signed distance. We first normalize the input mesh by translation and uniform scale, so that the geometry center sits at origin, and the maximum range of dimension of the normalized mesh's bounding box is ranging from -1 to 1. In order to represent the original geometry precisely, we hope that our target mesh has more than 10k triangles. Therefore, we set the length of a unit cube as 0.01 in the uniform grid.

For isosurface extraction, we only care about the distance value around the zero-isosurface. Therefore, instead of estimating the nearest distance between a cube and the input mesh, we can only record the sign of the cube, which includes positive (outside the surface), zero (intersecting the surface) or negative (inside the surface). We can extract the surfaces between the positive cubes and occupied cubes to represent the manifold. To further make the representation more efficient, we adopt the octree instead of the uniform grid. To achieve the same surface quality, we need the grid to have the same resolution around zero distance, but much coarser resolution in other volumes. As a result, the scale of the computation and storage will increase linearly with the scale of the final mesh.

\section{Volume Computation}
In this section, we describe the concrete method for building the octree we need.
\subsection{Build Octree}
We start building the octree from a single root node. For each node, we store the effective bounding box the node represents. Additionally, we store the set of triangles the bounding box contains or intersects. In order to fully cover the normalized triangle mesh, we make each dimension of the root's bounding box range from -1.1 to 1.1. As a result, it also stores all triangles in the normalized mesh. We define occupied node as those whose triangle set is non-empty.

Our target is to build an octree where each occupied node has the finest resolution (0.01). Therefore, we keep subdividing the occupied nodes into eight children unless they reach the final resolution, and reassign the triangle set of the node to its children. We do not subdivide the empty nodes which do not contain any triangle. A detailed description is shown in algorithm~\ref{alg:1}.
\begin{algorithm}
 \KwData{Input triangles $M$, bbox = BoundingBox($M$)}
 \KwResult{Octree T}
 T.bbox = bbox\;
 T.triangles = M\;
 \If {M is empty}{
    T.status = Empty\;
 } \Else {
    T.status = Occupied\;
    \For {i $\leftarrow 1\; to\; 8$} {
        SubBbox[i]$\leftarrow$ SubBoundingBox(bbox, i)\;
        
        SubM[i] $\leftarrow$ ContainedBy(M, SubBbox[i])\;
        
        T.children[i] = BuildOctree(SubM[i], SubBbox[i])\;
    }
 }
 \Return T\;
 \caption{BuildOctree}
 \label{alg:1}
\end{algorithm}

\subsection{Build Connection}
Because our surface extraction method relies on detection of pairs of neighboring occupied-empty nodes with the direction of the connection ($\pm x$,$\pm y$,$\pm z$), we need to build the connection between neighboring nodes. The challenge is that some nodes are larger than others, resulting in a multiple-to-multiple connection. Because all our occupied nodes are in the finest resolution, we only store the connection from occupied nodes to others.

We build the connections in the recursive way. We start from the root of the tree, and build connections inside each child node. For each neighboring pairs of children, we build connection between two nodes. Note that there are totally 12 pairs of neighboring children, considering there are 12 edges of a cube. If both nodes are empty, we don't record connection. If only one node is empty, we recursively build the connection between children of the other node and this empty node. If both nodes are occupied, we recursively connect the four neighboring children of these two nodes. The detailed description is shown in algorithm~\ref{alg:2}~\ref{alg:3}.

\begin{algorithm}
 \KwData{Input Octree T}
 \If{T.status = Empty} {
    \Return\;
 }
 \For {i $\leftarrow 1\; to\; 8$} {
    BuildConnection(T.children[i])\;
 }
 \For {i $\leftarrow 1\; to\; 12$} {
    ConnectNodes(T.children[Neighbor[i].x], T.children[Neighbor[i].y])\;
 }
 \caption{BuildConnection}
 \label{alg:2}
\end{algorithm}

\begin{algorithm}
  \KwData{Input two nodes $N_l$,$N_r$}
  \If {$N_l$ = Empty and $N_r$ = Empty} {
    \Return\;
  } 
  \If {$N_l$ = Empty} {
      $SubN_l = \{N_l,N_l,N_l,N_l\}\;$
  }
  \Else {
      $SubN_l = N_l.children\{s_1, s_2, s_3, s_4\}\;$
  }
  \If {$N_r$ = Empty} {
      $SubN_r = \{N_r,N_r,N_r,N_r\}\;$
  }
  \Else {
      $SubN_r = N_l.children\{s_5, s_6, s_7, s_8\}\;$
  }
  
 \For {i $\leftarrow 1\; to\; 4$} {
    ConnectNodes($SubN_l[i]$,$SubN_r[i]$)\;
 }
 \caption{ConnectNodes}
 \label{alg:3}
\end{algorithm}
\section{Manifold Surface Extraction}
We generate the isosurface as the faces between positive cubes and occupied cubes, which is similar to the idea of marching cubes~\cite{lorensen1987marching}. However, it is tricky to define whether the space is positive (outside) or negative (inside) the surface, especially if the surface contains holes. In our definition, if there is a path from a certain cube to the boundary which is not occupied by any triangle, the cube is positive. We initialize all boundary cubes as positive, and apply a BFS algorithm to expand the positive cubes. All remaining cubes which are not occupied are negative. To search for all these faces, we recursively visit all occupied nodes and their neighbors to detect occupied-positive connections.

\begin{figure}
    \centering
    \includegraphics[width=0.48\linewidth,height=0.28\linewidth]{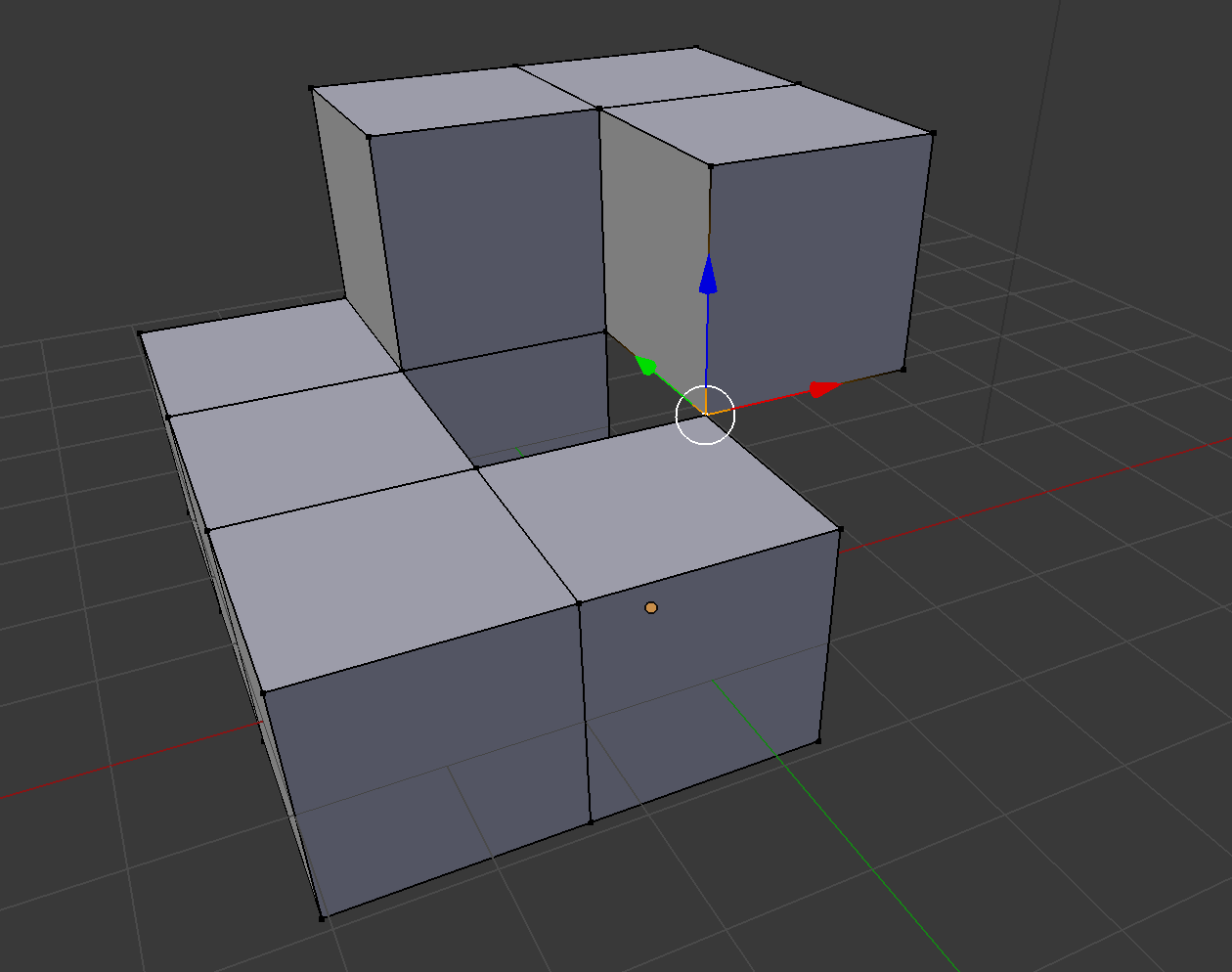}
    \includegraphics[width=0.48\linewidth,height=0.28\linewidth]{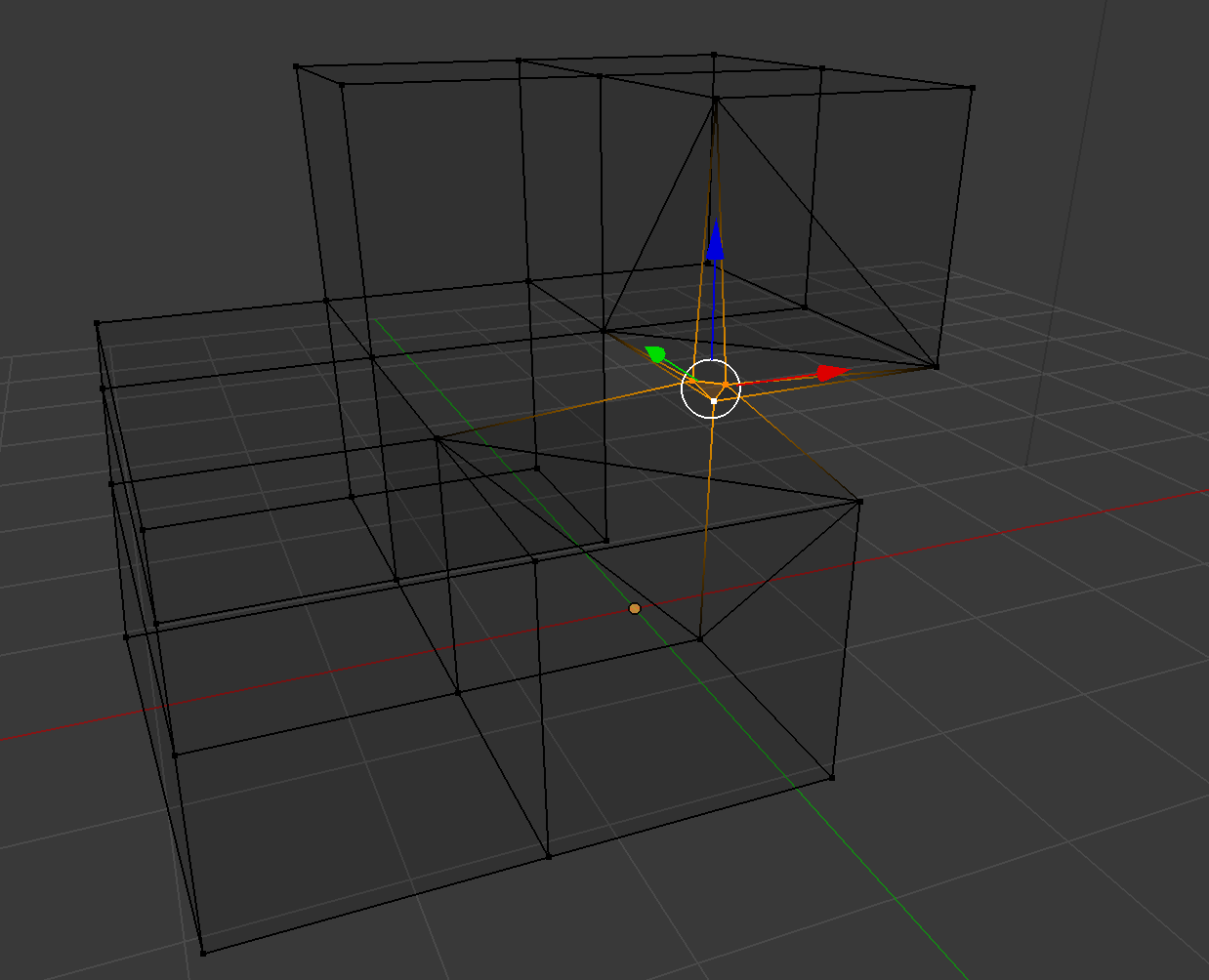}
    \includegraphics[width=0.48\linewidth,height=0.28\linewidth]{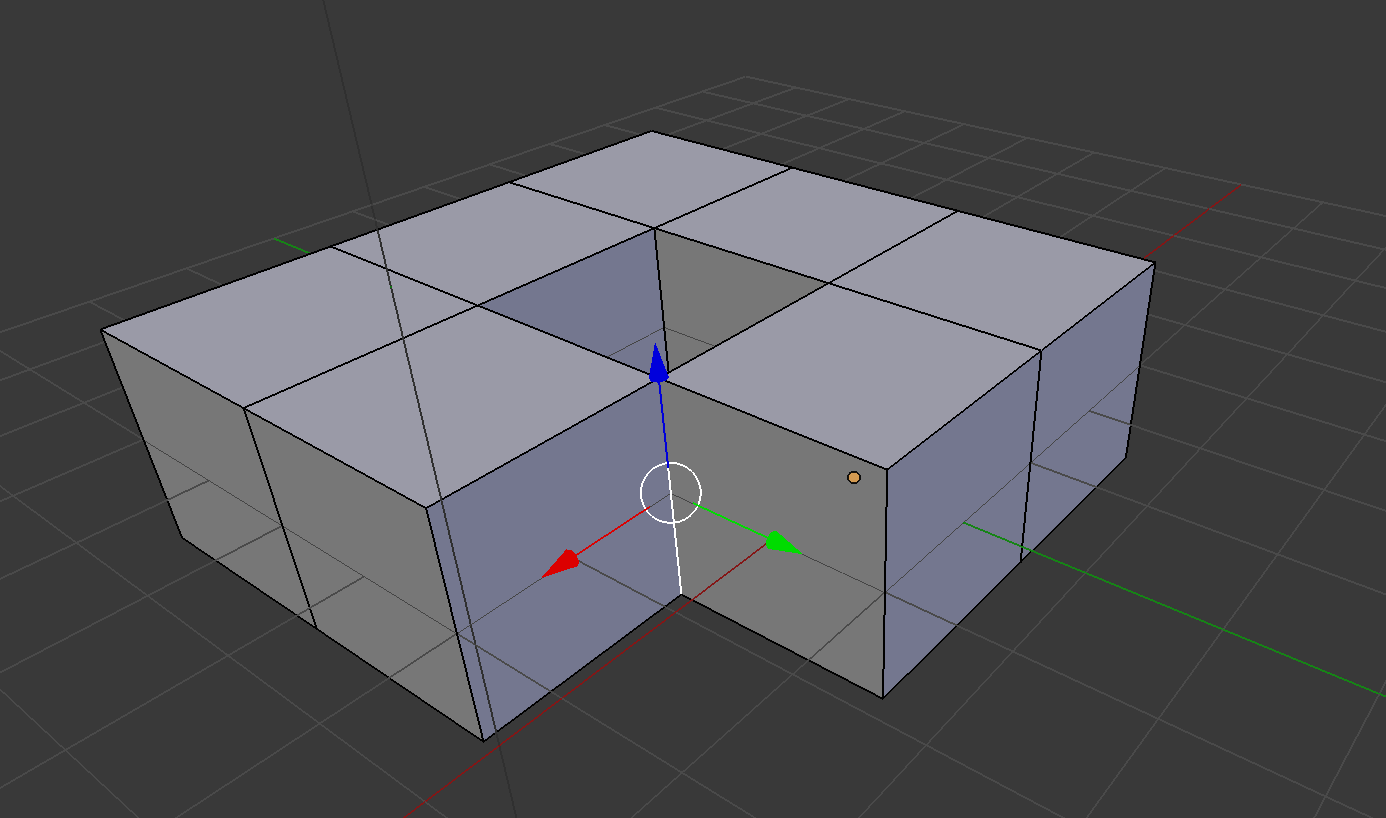}
    \includegraphics[width=0.48\linewidth,height=0.28\linewidth]{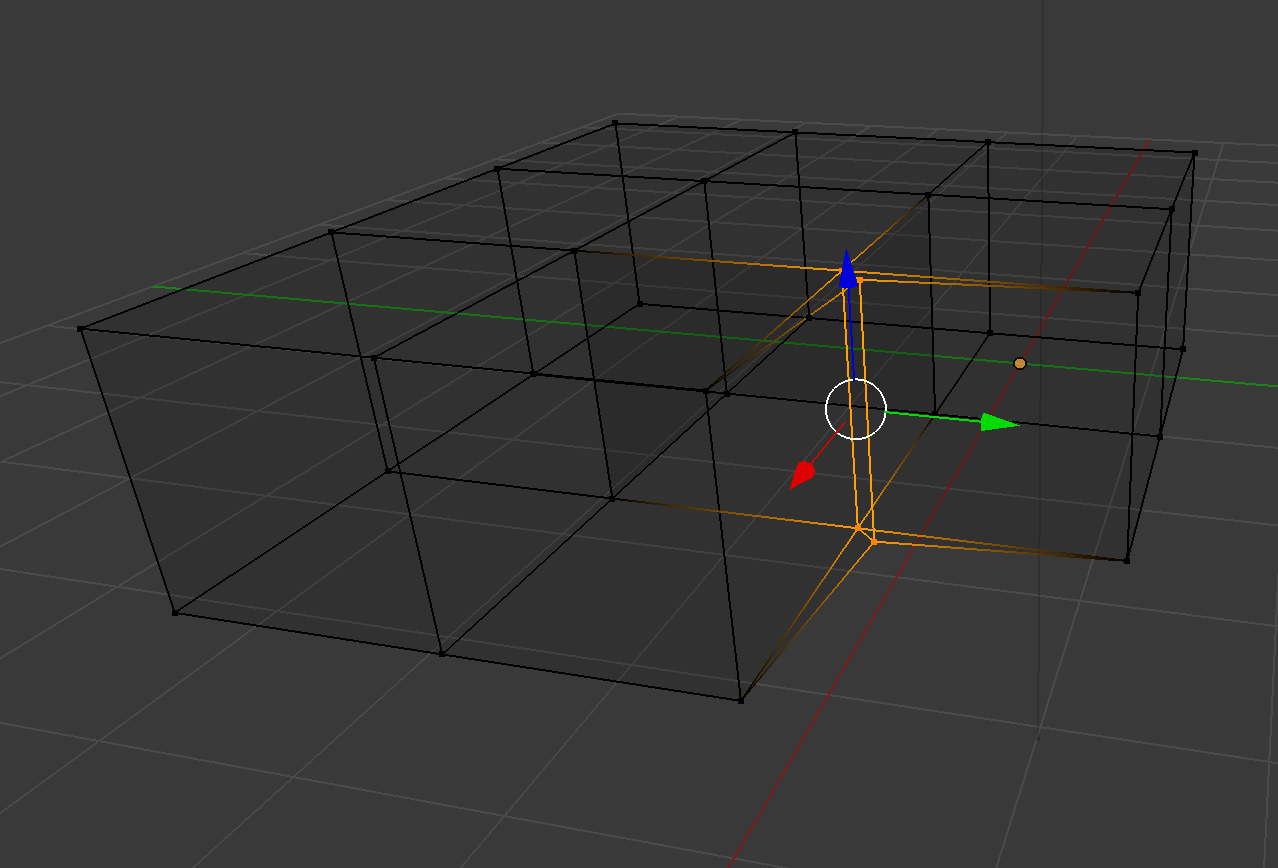}
    \caption{Non-manifold Examples}
    \label{fig:non-manifold}
\end{figure}

However, this generation method is not guaranteed to form a 2-Manifold. The two cases are shown in figure~\ref{fig:non-manifold}. In the first row (left), though all edges are connected with only two faces, two cubes meets at a single vertex (selected), this leads to the ambiguity whether the volume of the two cubes are connected through the vertex. In the second case (left), two cubes share a common edge, which connects 4 faces. We remove this ambiguity splitting the vertex into three or the edge into two, as shown in the second column, assuming volumes are always connected.

\section{Projection}
The representation of the cube faces as the manifold is not precise, because the vertices are not always sitting at the original triangle mesh. The way to fix it is to project the vertices onto the given mesh. Because we already build the relation between occupied cubes and triangles they contain, it is efficient to compute the nearest triangle to a certain vertex of the extracted face. We iteratively move all extracted vertices closer to their nearest triangles along the direction parallel to their normals with a fixed small step. A laplacian smooth is applied after each step to prevent face flip.

\section{Results}
We apply our algorithm to all ShapeNet models, which contains hole and thin structures. As a result, all our conversion are topologically 2-Manifolds. We additionally check whether there are triangle flips in the model. As a result, only about 5\% contains face flip, mostly caused by thin face structures in the original mesh.
\bibliographystyle{acmsiggraph}
\bibliography{template}
\end{document}